# Spatial phase sensitivity for oscillators close to the saddle-node homoclinic bifurcation

Jinjie Zhu[*]

*School of Mechanical Engineering, Nanjing University of Science and Technology, Nanjing 210094, China*

## Abstract

The traditional phase sensitivity function (PSF) has manifested its efficacy in investigating synchronization behaviors for limit-cycle oscillators. However, some subtle details may be ignored when the phase value is accumulated in space or the perturbation is space-dependent. In this paper, we compared spatial PSF with the traditional PSF for oscillators close to the saddle-node homoclinic (SNH) bifurcation, also known as saddle-node on invariant circle (SNIC) bifurcation. It is found that the spatial phase sensitivity function could reveal the phase accumulation feature on the limit cycle. Moreover, it is proved that for any two-dimensional smooth dynamical system, type II phase response curve is the only possible type. Finally, the synchronization distributions of uncoupled SNH oscillator driven by common and independent noises are studied, which shows the space-dependent coupling function of common noise could have significant influences on the synchronization behavior.

## 1. Introduction

Phase reduction approach [1] is an efficient method to investigate the dynamical behaviors of periodic oscillators subjected to weak perturbations by utilizing the phase variable to reduce the original system into one dimension. As the key part, the phase sensitivity function (PSF) [2] or the infinitesimal phase response curve (PRC) [3] directly reflects the impact of the perturbation on the phase change (advance or delay). It has a deep connection with the conception of isochrons [4,5] and Floquet eigenvectors [2]. Due to its simplicity and efficiency, this approach has been widely employed in coupled or uncoupled oscillators, delayed systems, quantum synchronization, network dynamics, reaction-diffusion systems, hybrid systems and relaxation oscillators [6–15], etc.

In neuroscience, Hansel *et al.* [16] first found two types of PRC. Type I will always advance the phase by excitatory postsynaptic potentials while Type II can either delay or advance the phase. Later, Ermentrout [17] related the types of PRC with the classification of the excitable membranes by Hodgkin and further found Type II is more favorable for stochastic synchronization [18]. Brown *et al.* [19] made a probabilistic analysis of phase response for four kinds of neuron models encompassing four generic bifurcations and obtained the scaling behaviors near the bifurcations, which contains both types of the PRC.

[*] Corresponding author.
 *E-mail:* zhujinjie95@njust.edu.cn

As a representative model of Type I PRC, oscillators close to saddle-node homoclinic bifurcation (SNH oscillator for short in this paper) have been analyzed by the phase reduction approach [3,17,19]. Different from the Type II PRC, the phase of the SNH oscillator can accumulate around the bifurcation point. But this could not be unveiled by the phase response curve or the phase sensitivity function, as the PSF is measured on the phase not space. Therefore, to exhibit the spatial accumulation characteristics, we will compare the traditional and the spatial PSF in this paper. The mathematical model and its bifurcation behavior will be discussed in Section 2. The traditional and spatial PSF will be analyzed in detail in Section 3. In Section 4, we study the noise-induced synchronization of the uncoupled SNH oscillator for constant and space-dependent coupling functions of common noise. Finally, the discussions and conclusions are given in Section 5.

## 2. Mathematical model and setup

To obtain generic results, we consider the canonical oscillator close to the saddle-node homoclinic bifurcation given by Yuri A. Kuznetsov [20]:

$$\dot{x} = x(1 - x^2 - y^2) - y(1 + \alpha + x),$$
$$\dot{y} = x(1 + \alpha + x) + y(1 - x^2 - y^2).$$
(1)

Under the polar transformation $x = r\cos(\varphi), y = r\sin(\varphi)$, system (1) can be easily transformed as follows:

$$\dot{r} = r(1 - r^2),$$
$$\dot{\varphi} = 1 + \alpha + r\cos(\varphi).$$
(2)

where $r$ represents the radius and $\varphi$ is the angle variable (in this paper, we use angle instead of phase to avoid confusion with the phase of the oscillation). The parameter $\alpha$ is the bifurcation parameter. The system will always have an unstable equilibrium (0, 0) regardless of the value of $\alpha$. For $-2 < \alpha < 0$, the system will have two other equilibrium points: $r$=1, $\varphi = \pm\cos^{-1}(-(1+\alpha))$ (black and white dots in Fig. 1). For $\alpha = 0$, the system will have a saddle-node point $r$=1, $\varphi = \pi$ and undergo the saddle-node homoclinic bifurcation. For $\alpha > 0$, a stable limit cycle appears. It should be noted that the circle with radius $r$=1 is an invariant cycle for all values of $\alpha$. The bifurcation scenario is illustrated in Fig. 1 for three typical parameters of $\alpha$.

To have a limit cycle, we set $\alpha > 0$ in the following and investigate cases when $\alpha$ is close to the bifurcation value $\alpha = 0$. For simplicity, we name the oscillator close to the saddle-node homoclinic bifurcation as the SNH oscillator. It can be readily checked that the $x$-nullcline is always pinned at (±1, 0) and the $y$-nullcline is always pinned at (0, ±1). This property will be useful for determining the positive or negative phase response at these points later. The angle frequency of the limit cycle is $\omega=\sqrt{\alpha(\alpha+1)}$, and the period is $2\pi/\omega$. It can be seen that as $\alpha \to 0$, the period will approach infinity. The time-angle curve can also be readily obtained from system (2) as:

$$\varphi(t) = 2\arctan\left(\frac{\omega}{\alpha}\tan\left(\frac{\omega t}{2}\right)\right) \tag{3}$$

The time-angle curves for system (2) are illustrated in Fig. 2 for different bifurcation parameter $\alpha$. The period of the oscillator gradually increases as $\alpha \to 0$ and the major time course of the trajectory will remain close to angle $\varphi = \pi$ which is consistent with the saddle-node point when $\alpha = 0$.

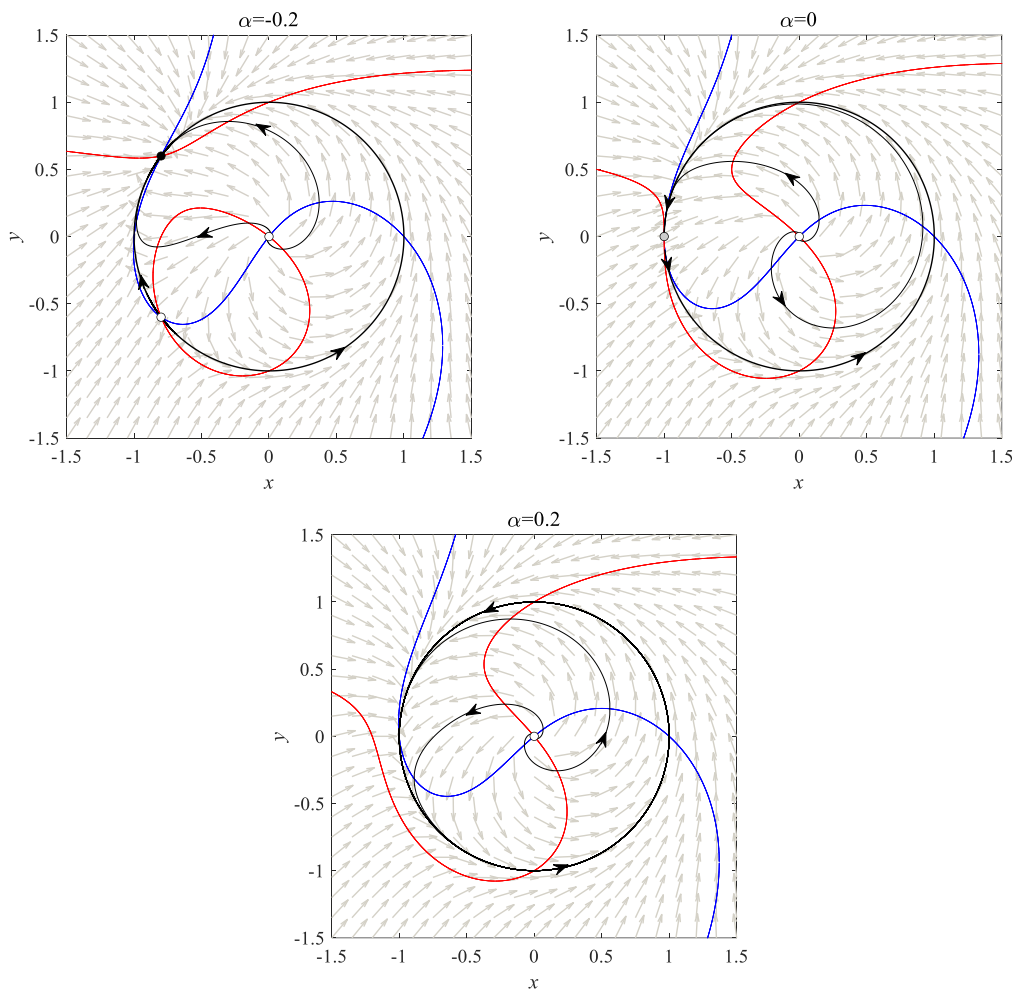

Fig. 1. Bifurcation scenario for the saddle-node homoclinic bifurcation of system (1). The black dot represents the stable equilibrium and the white dots denote the unstable ones. The grey dot for $\alpha=0$ is the saddle-node point. The blue and red curves are the *x*- and *y*-nullclines, respectively. The grey arrows denote the direction of the vector field.

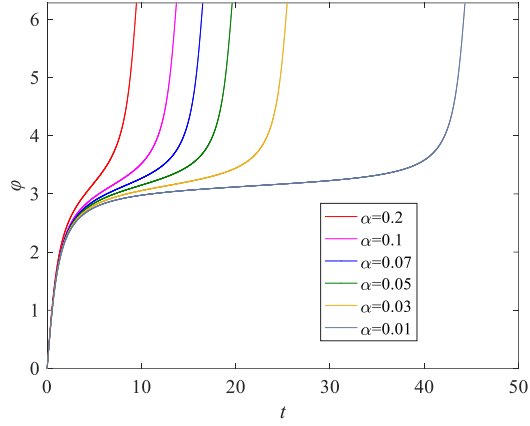

Fig. 2. The time-angle curves for system (2) for different bifurcation parameter $\alpha$. Only one cycle is plotted.

## 3. Phase sensitivity function

*3.1. Traditional phase sensitivity function (tPSF)*

Considering the weakly perturbed situation, to simplify the notation, we rewrite system (1) as:

$$\dot{\mathbf{X}} = \mathbf{F}(\mathbf{X}) + \mathbf{P}(\mathbf{X},t) \qquad (4)$$

where $\mathbf{F}(\mathbf{X})$ is the original vector field and $\mathbf{P}(\mathbf{X},t)$ is the weak perturbation. By the standard phase reduction approach [1,21], it can be reduced to the system with a single phase variable:

$$\frac{\mathrm{d}}{\mathrm{d}t}\theta(t) = \frac{\partial \theta}{\partial \mathbf{X}} \cdot \frac{\mathrm{d}\mathbf{X}}{\mathrm{d}t} = \frac{\partial \theta}{\partial \mathbf{X}} \cdot \left(\mathbf{F}(\mathbf{X}) + \mathbf{P}(\mathbf{X},t)\right) = \omega + \frac{\partial \theta}{\partial \mathbf{X}} \cdot \mathbf{P}(\mathbf{X},t) \qquad (5)$$

where $\theta(t)$ is the phase variable, which by definition linearly grows with time $t$ as $\theta(t) = \omega t$. Because the perturbation is assumed weak, equation (5) can be further reduced, to the linear order of the perturbation as:

$$\frac{\mathrm{d}}{\mathrm{d}t}\theta(t) \approx \omega + \left.\frac{\partial \theta}{\partial \mathbf{X}}\right|_{\mathbf{X}=\mathbf{X}_0(t)} \cdot \mathbf{P}(\mathbf{X}_0(t),t) = \omega + \mathbf{Z}(\theta) \cdot \mathbf{P}(\mathbf{X}_0(t),t) \qquad (6)$$

where $\mathbf{X}_0(t)$ is the limit cycle of the unperturbed system and $\mathbf{Z}(\theta) = [Z_x(\theta), Z_y(\theta)]$ is the so-called phase sensitivity function (PSF) or the infinitesimal phase response curve (PRC), which is the gradient of the phase variable on the limit cycle [1–3]. Through equation (6), it is obvious that if the PSF can be obtained, then the phase dynamics can be captured accordingly.

Although PSF cannot be analytically calculated in most cases, there are a handful of numerical methods. Here we apply two methods, namely, the direct method and the adjoint method [1], to calculate the PSF for the SNH oscillator. Figure 3 illustrates the results for $\alpha = 0.2$ by these two methods and they agree with each other within acceptable tolerance.

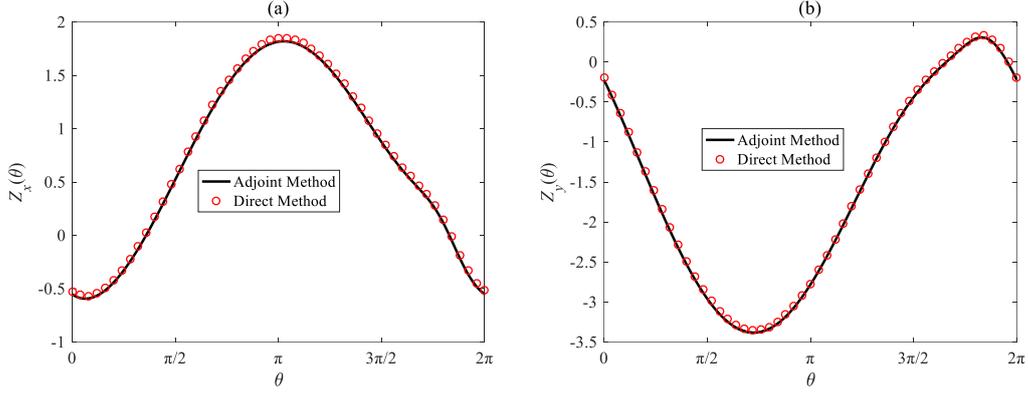

Fig. 3. Phase sensitivity function for SNH oscillator via the direct method and the adjoint method. (a) The *x* component; (b) The *y* component. The parameter $\alpha = 0.2$. For the direct method, the perturbation strength is fixed as 0.005. The initial phase $\theta=0$ is chosen as (0, 1).

For $\alpha \to 0$, the results for PSF have been investigated widely [3,17,19]. However, for comparison, we still calculated them by adjoint method as in Fig. 4. By decreasing $\alpha$, the PSF gradually approaches to the curve $A\sin^2(\theta)$ or $K(1-\cos(2\theta))$ [3,17], where *A* or *K* are proper constants. The limit situation shows that the PSF along *x*-direction, i.e. $Z_x(\theta)$, will be nonnegative (similarly, nonpositive for $Z_y(\theta)$). This is called the Type I PRC by Ermentrout. However, here we would like to comment that for a smooth two-dimensional dynamical system, it's impossible for the PSF or PRC to be strictly nonnegative. The nonnegative PSF will only be possible for the limit situation, e.g., by making $\alpha = 0$ in SNH oscillator (but this will cause the period of the limit cycle to be infinity) or for non-smooth or discontinuous system (e.g., the quadratic integrate-and-fire neuron [3]). We will discuss this point in the next subsection for the spatial phase sensitivity function.

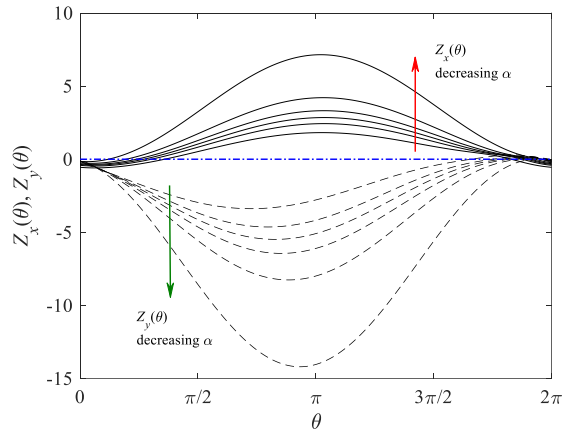

Fig. 4. Phase sensitivity function for SNH oscillator for different bifurcation parameter $\alpha$. The solid curves and the dashed curves represent results for $Z_x(\theta)$ and $Z_y(\theta)$, respectively. The red and green arrows show the change of the PSF by decreasing $\alpha$. The blue dash-dotted line is the zero-value line. Parameters are: $\alpha = 0.2, 0.1, 0.07, 0.05, 0.03, 0.01$.

## 3.2. Spatial phase sensitivity function (sPSF)

The traditional phase sensitivity function or tPSF for SNH oscillator shows an overall increase in amplitude when the parameter is close to the bifurcation (see Fig. 4). And the nonnegative interval for $Z_x(\theta)$ extends gradually to the whole period. However, at the same time, as far as the space is concerned, the phase is compressed around the angle $\varphi = \pi$. Figure 5 shows this feature by depicting the phase values on the limit cycle. Although the phase by definition uniformly increases with time, it is not uniform in space. As $\alpha \to 0$, the phase accumulates around the phase $\theta = \pi$ or the space position (-1, 0).

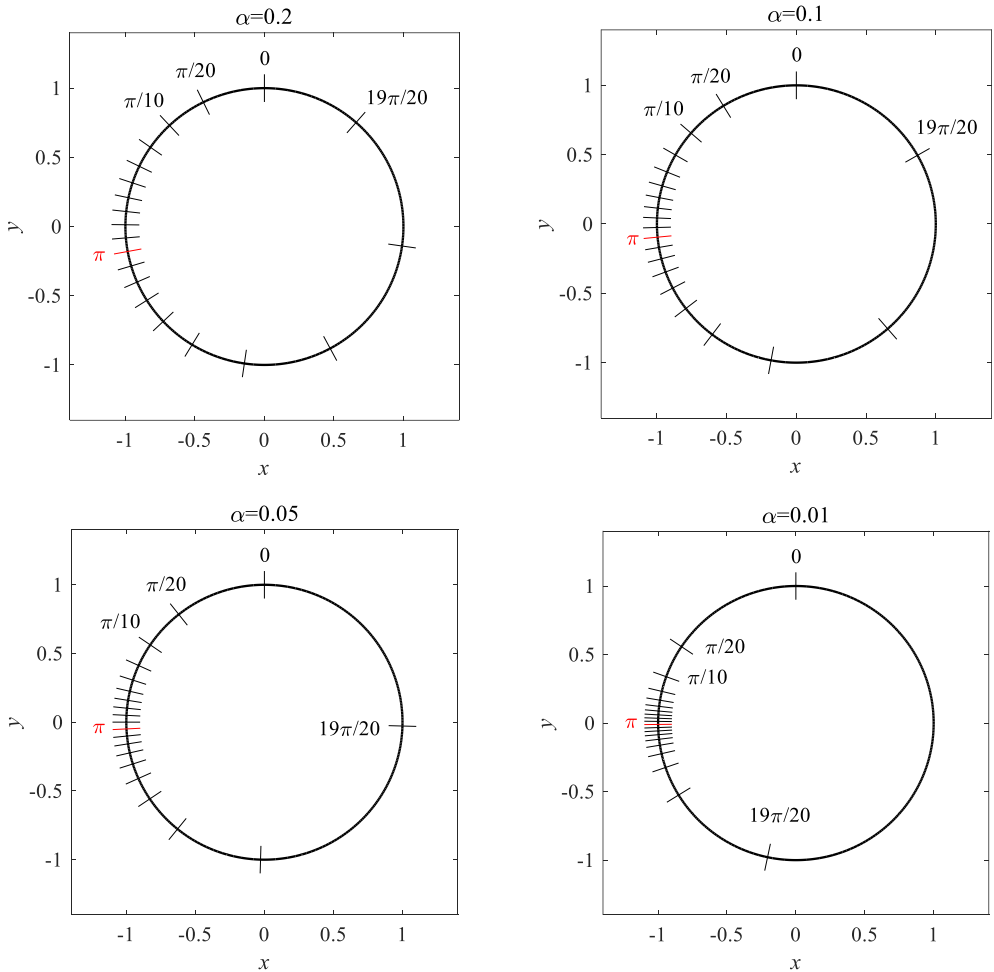

Fig. 5. The phase value $\theta$ on the limit cycle for different bifurcation parameters. The phase is equally distributed along the limit cycle with the interval $\pi/20$.

To illustrate the phase sensitivity to the space position, we calculate the phase sensitivity as a function of space position along the limit cycle. Similar to the traditional phase sensitivity function, we may define a space variable $\varphi$ which varies from 0 to $2\pi$. By assumption, $\varphi$ will be uniform along the limit cycle within which $\varphi$ will increase linearly as the arc length (i.e. space). Because the limit cycle in our case is just the unit circle centered at the origin, and by virtue of Eq. (3) (note that the starting angle of Eq. (3) is 0 but the starting angle in Fig. 5 is $\pi/2$), the phase sensitivity as a function of space ($\varphi$ in our case) can be readily obtained. We will call it the spatial phase sensitivity function (sPSF). The sPSF is plotted in Fig. 6, where the parameter $\alpha$ is chosen the same as in Fig. 4. By comparing the sPSF with the tPSF in Fig. 4, it is obvious to note that the phase sensitivity is broad in time but narrow in space. The sPSF is peaked mainly around the angle $\varphi=\pi$ which is even more apparent as $\alpha$ decreases. This is consistent with result in Fig. 5 where the phase value accumulates at $\varphi=\pi$, so the phase change will be more sensitive to the perturbation there. Next, we want to obtain some quantitative or qualitative characteristics of the spatial phase sensitivity function.

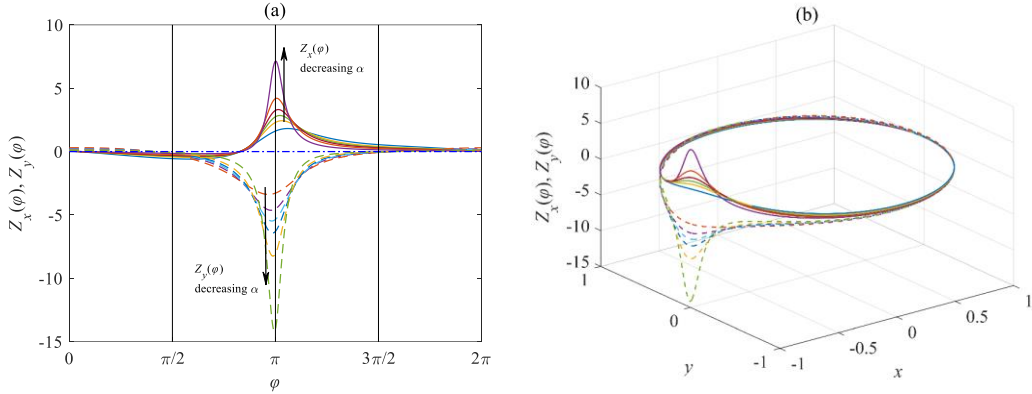

Fig. 6. The spatial phase sensitivity function (sPSF) for SNH oscillator for different bifurcation parameter $\alpha$. (a) The solid curves and the dashed curves represent results for $Z_x(\varphi)$ and $Z_y(\varphi)$, respectively. (b) The three-dimensional view for sPSF. Parameters are the same as in Fig. 4.

Considering the definition of the phase variable and the phase sensitivity function, the following condition always satisfies:

$$\mathbf{Z}(\mathbf{X}) \cdot \mathbf{F}(\mathbf{X}) = \omega \qquad (7)$$

which is also the normalization condition for the adjoint equation [1,17]. It is the reason for the origin (0, 0) to be the phaseless [3,5] equilibrium, as the vector field vanishes there. For SNH oscillator, as we mentioned previously, there are four intersection points for the limit cycle and the nullclines: ($\pm 1$, 0) for the x-nullcline and (0, $\pm 1$) for the y-nullcline. Take (1, 0) as an example. Because the x-component of the vector field $F_x(\mathbf{X})$ is zero and Eq. (7) goes as $Z_y(\mathbf{X}) \cdot F_y(\mathbf{X}) = \omega$, $Z_y(\varphi=0) = Z_y(\mathbf{X}(\varphi=0)) = \omega/F_y(\mathbf{X}(\varphi=0))$ would be positive because $F_y(\mathbf{X}(\varphi=0)) > 0$. Similarly, $Z_y(\varphi=\pi) < 0$, $Z_x(\varphi=\pi/2) < 0$ and $Z_x(\varphi=3\pi/2) > 0$ (It can be verified in Fig. 6(a) although the amplitude may be small).

To be more generic, for any smooth two-dimensional dynamical system, and for perturbation on any direction, say $\vec{e}_i$, the limit cycle will have at least two positions. For one position, the vector field on it will be along $\vec{e}_i$ and the other will be against $\vec{e}_i$ (or along the direction $-\vec{e}_i$). Thus by Eq. (7), the sPSF along $\vec{e}_i$ will be positive at the former position while negative at the latter position. Therefore, for any smooth two-dimensional dynamical system, strictly nonnegative PSF (or Type I PRC) is impossible; there are at least two positions have PSF with opposite signs (and their neighboring area by continuity) although the amplitude may be small. In other words, every smooth two-dimensional dynamical system possesses Type II PRC. Figure 7 illustrates the schematic diagram for the explanation. For systems with higher dimension, it remains to validate whether there is a similar result as the two-dimension case.

In the following section, we will show that sPSF can reveal the importance of space-dependent perturbation on the system.

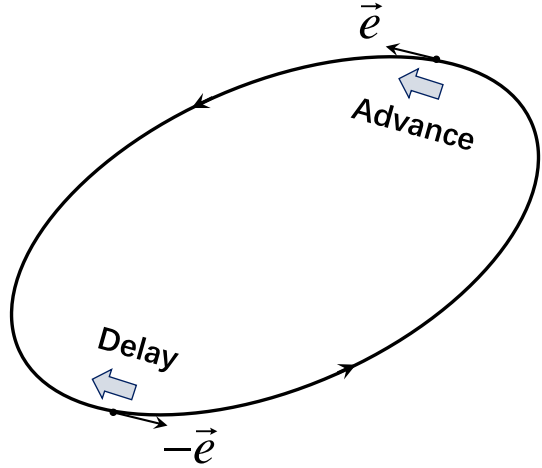

Fig. 7. Schematic diagram of positions for phase advance and delay.

## 4. Noise-induced synchronization of uncoupled SNH oscillators

In this section, we consider uncoupled SNH oscillators driven by common and independent noises. The governing equation is as follows:

$$\dot{\mathbf{X}}^i = \mathbf{F}(\mathbf{X}^i) + \sqrt{D_1}\mathbf{G}(\mathbf{X}^i)\xi(t) + \sqrt{D_2}\mathbf{H}(\mathbf{X}^i)\mathbf{v}^i(t) \qquad (8)$$

where $\xi(t)$ is the common noise shared by all the SNH oscillators and $\mathbf{v}^i(t)$ is the independent noise. They are assumed to be independent, identically distributed zero-mean Gaussian white noise and satisfy the correlation functions $\langle \xi_u(t)\xi_v(\tau)\rangle = \delta_{u,v}\delta(t-\tau)$, $\langle \mathbf{v}^i_u(t)\mathbf{v}^j_v(\tau)\rangle = \delta_{i,j}\delta_{u,v}\delta(t-\tau)$ and $\langle \xi_u(t)\mathbf{v}^i_v(\tau)\rangle = 0$. $D_1$ and $D_2$ denotes their strengths. According to Ref. [10], derived by Nakao *et al*., the stationary probability density distribution (PDF) of the phase difference

$\phi = \theta^i - \theta^j$ is given as:

$$U(\phi) = \frac{u_0}{D_1[g(0) - g(\phi)] + D_2 h(0)} \quad (9)$$

where $u_0$ is the normalization constant. The functions $g(\phi)$ and $h(0)$ are given as:

$$g(\phi) = \frac{1}{2\pi} \int_{-\pi}^{\pi} \mathbf{Z}(\theta)\mathbf{G}(\theta)\mathbf{G}^{\mathrm{T}}(\theta+\phi)\mathbf{Z}^{\mathrm{T}}(\theta+\phi)\mathrm{d}\theta \quad (10)$$

$$h(0) = \frac{1}{2\pi} \int_{-\pi}^{\pi} \mathbf{Z}(\theta)\mathbf{H}(\theta)\mathbf{H}^{\mathrm{T}}(\theta)\mathbf{Z}^{\mathrm{T}}(\theta)\mathrm{d}\theta \quad (11)$$

For $D_1 = 0, D_2 \neq 0$, which is the case of independent noise alone, via Eq. (9), the PDF of phase difference will be a uniform distribution: $U(\phi) = 1/2\pi$. For $D_1 \neq 0, D_2 = 0$, which is the case of common noise alone, the PDF will be a delta function at zero phase difference. This is the phenomenon of common noise-induced complete synchronization. In this paper, we consider the existence of both kinds of noises.

We set noise strength as $D_1 = 0.005, D_2 = 0.0001$. Figure 8 illustrates the PDF of phase difference by Eq. (9) and by Monte Carlo simulation for different coupling function $\mathbf{G}(\mathbf{X}^{(i)})$ of common noise. For Fig. 8(a), $\mathbf{G}(\mathbf{X}^{(i)}) = \mathbf{G}_1 = \mathrm{diag}(1,1)$, the common noise is uniform in space. The theoretical and numerical results show that the PDF of the phase difference is peaked at $\phi = 0$. This is the typical one-cluster synchronization. The inset in Fig. 8(a) shows the spatial distribution of the oscillators around the limit cycle. For space-dependent coupling function, we analyze three kinds of exponential kernel, which are: (1) $\mathbf{G}(\mathbf{X}^{(i)}) = \mathbf{G}_2 = e^{-((x+1)^2+y^2)} \cdot \mathrm{diag}(1,1)$; (2) $\mathbf{G}(\mathbf{X}^{(i)}) = \mathbf{G}_3 = e^{-(x^2+y^2)} \cdot \mathrm{diag}(1,1)$; (3) $\mathbf{G}(\mathbf{X}^{(i)}) = \mathbf{G}_4 = e^{-((x-1)^2+y^2)} \cdot \mathrm{diag}(1,1)$. They are peaked at (-1, 0), (0, 0), (1, 0), respectively. The results are displayed in Fig. 8(b)-7(d). The PDFs for Fig. 8(a) and 7(b) are almost the same. This shows that the constant coupling function $\mathbf{G}(\mathbf{X}^{(i)})$ and the one with the exponential kernel centered at (-1, 0) give rise to similar synchronization behavior for the SNH oscillators. However, for exponential kernel centered far from (-1, 0), the PDF could be much flatter which shows a rather weaker synchronization (see Fig. 8(c)) and could even induce no synchronization with a nearly uniform distribution (see Fig. 8(d)).

The above observation cannot be unveiled by the traditional phase sensitivity function but can be simply inferred from the spatial phase sensitivity function. As we can see from the sPSF in Fig. 6. The sPSF is peaked around $\varphi = \pi$ or (-1, 0). Therefore, perturbation around this position can have a more significant influence on the phase change thus the synchronization behaviors.

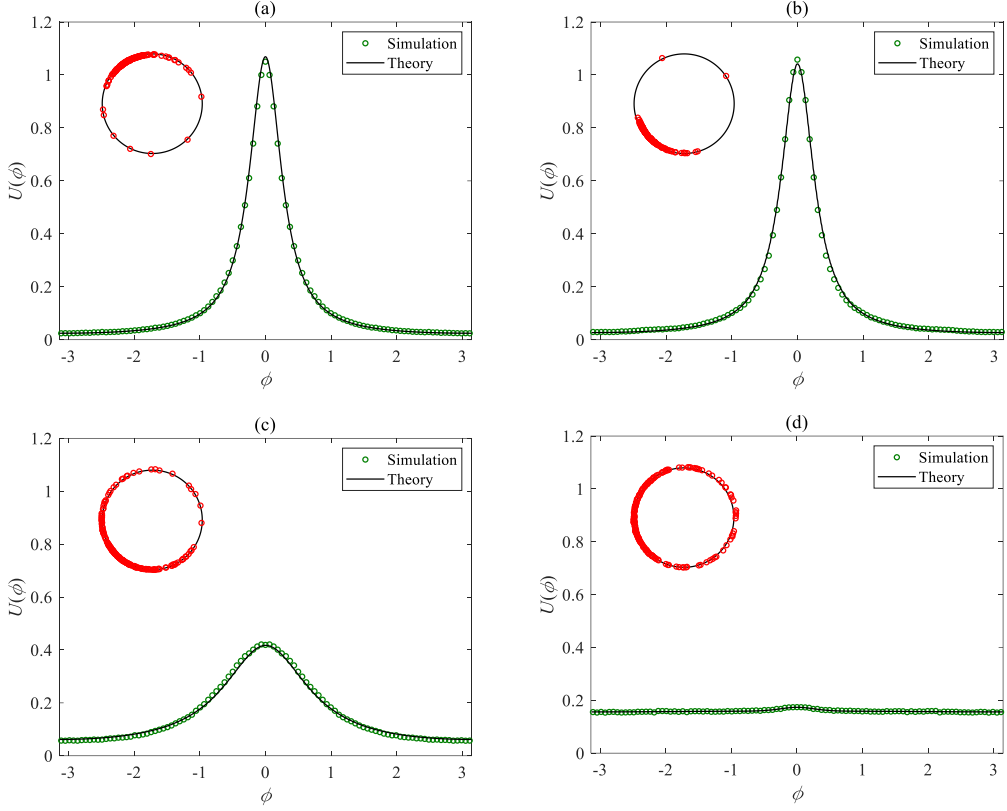

Fig. 8. The stationary probability density function of the phase difference for the SNH oscillators for different coupling function $\mathbf{G}(\mathbf{X}^{(i)})$ of common noise. The coupling functions $\mathbf{G}(\mathbf{X}^{(i)})$ are: (a) $\mathbf{G}(\mathbf{X}^{(i)}) = \text{diag}(1,1)$; (b) $\mathbf{G}(\mathbf{X}^{(i)}) = e^{-((x+1)^2+y^2)} \cdot \text{diag}(1,1)$; (c) $\mathbf{G}(\mathbf{X}^{(i)}) = e^{-(x^2+y^2)} \cdot \text{diag}(1,1)$; (d) $\mathbf{G}(\mathbf{X}^{(i)}) = e^{-((x-1)^2+y^2)} \cdot \text{diag}(1,1)$; The coupling function for independent noise is fixed as $\mathbf{H}(\mathbf{X}^{(i)}) = \text{diag}(1,1)$. Other parameters are: $\alpha = 0.2$, $D_1 = 0.005$, $D_2 = 0.0001$. The theoretical results are obtained by Eq. (9) and the numerical results are obtained by Monte Carlo simulation via Euler method. The insets display typical instantaneous distributions on the limit cycle.

For situations of other values of the bifurcation parameter $\alpha$, we calculate the PDFs in Fig. 9 for the same four kinds of coupling function of common noise. The results show a similar tendency (only slight difference is observed such as the insets in Fig. 9).

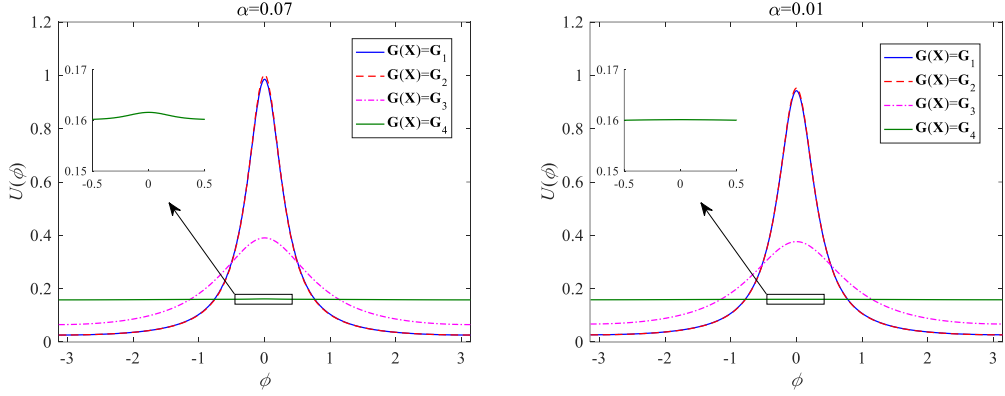

Fig. 9. The stationary probability density function of the phase difference for different bifurcation parameters. The four kinds of the coupling function are the same as in Fig. 8. The results are obtained by Eq. (9). The insets display the local enlargement within the black rectangle.

It should be noted that although the snapshots of instantaneous distributions of the SNH oscillators can display, to some extent, the synchronization behavior of the ensembles, it may cause some illusions only based on them. Figure 10 compares the snapshots of instantaneous distribution of oscillators on the limit cycle for different parameters and coupling functions. From the snapshots, for each value of $\alpha$, the degree of synchronization gradually attenuates from left to right as expected. However, for the same coupling function, the instantaneous distributions of the SNH oscillators show a more concentrated view for smaller $\alpha$ (from top to bottom in Fig. 10), which doesn't assure a more synchronized state. In fact, the PDF demonstrates that as $\alpha$ decreases, the peaks reduce slightly (although not obvious), which leads to an opposite conclusion. It can be explained by the spatial nonuniform distribution of the phase value as is revealed previously in Fig. 5. This is quite different from the uniform situation where snapshots of instantaneous distribution of the oscillators can clearly perform their clustering features as in Ref. [10] for the Stuart-Landau oscillators. Another important point could be observed is that as $\alpha \to 0$, the snapshots of the oscillator distribution will change less compared with the stationary PDF as is demonstrated on the bottom of Fig. 10 for $\alpha=0.01$ (there is accumulation of oscillators even when the PDF is almost uniform). This is also caused by the phase accumulation. In that case, to better exhibit the synchronization, the limit cycle could be replaced by the phase circle where the phase is uniformly distributed on the circle.

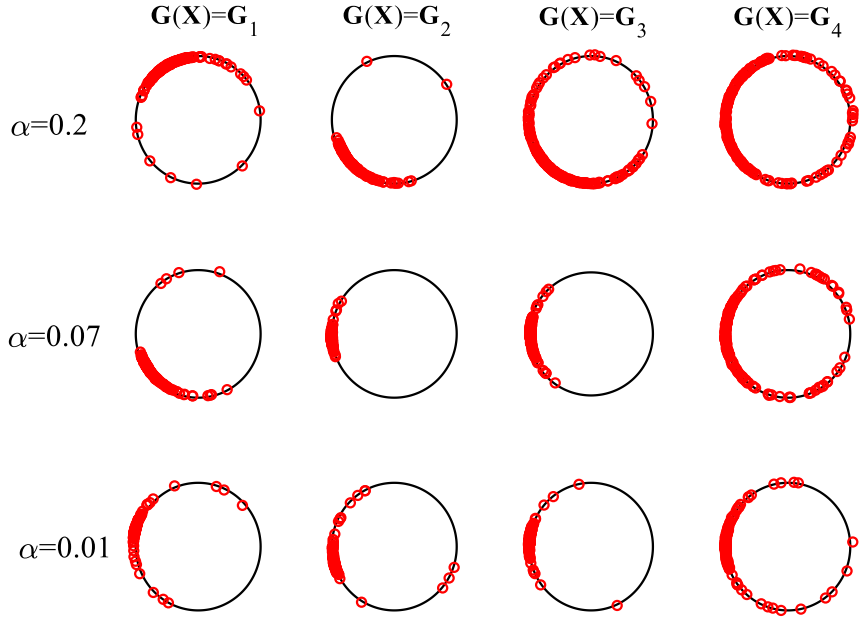

Fig. 10. Snapshots of instantaneous distribution of oscillators on the limit cycle for different parameters and coupling functions.

## 5. Discussions and conclusions

In summary, the spatial and traditional phase sensitivity of oscillator close to saddle-node homoclinic bifurcation are investigated in this paper. It is found that for the SNH oscillator close to the bifurcation, the traditional phase sensitivity function (tPSF) will be close to the quadratic sine curve which is a well-known result as the Type I PRC. However, the spatial phase sensitivity function (sPSF) indicates that there will always be positions in space for PSF to be positive and negative. The nonnegative PRC can only be realized in the limit situation or for non-smooth or discontinuous dynamical systems. That means for any smooth two-dimensional dynamical system, there could only be Type II PRC with phase sensitivity function having both positive and negative values. Due to the accumulation of the phase value around the angle $\varphi=\pi$ or the position (-1, 0), the sPSF will correspondingly accumulate there. For uncoupled SNH oscillators driven by common and independent Gaussian white noises, the result shows that space-dependent coupling function of common noise could have significant influences on the synchronization behavior of the SNH oscillators. Similar results can be obtained among other bifurcation parameters revealed by the stationary probability density function. This could also be demonstrated by the snapshots of instantaneous distribution of the oscillators on the limit cycle. However, attention should be paid that the snapshots may give rise to wrong assertion of the degree of synchronization because of the accumulation of the phase value. This is consistent with a recent work by Freitas *et al.* [22] wherein they found that topological properties of an attractor have no bearing on phase coherence. It is suggested to utilize the uniformly distributed phase circle instead of the limit cycle in that case.

The phase sensitivity function is powerful in analyzing synchronization induced by rhythmic perturbation or other time correlated excitation. When the perturbation is space correlated (e.g., the space-dependent control) or the phase is extremely locally accumulated, the spatial phase sensitivity function may be more intuitive to the understanding for the dynamical behaviors of limit-cycle oscillators. Besides the SNH oscillator, it deserves efforts to uncover other phenomena by sPSF for different bifurcations which render limit-cycle oscillations. The space-dependent perturbation is restricted to common noise in this paper. So, it would be interesting to investigate other types of spatial perturbations or even spatiotemporal coupling perturbations in future works.

## Acknowledgements

This study was funded by Natural Science Foundation of Jiangsu Province of China (BK20190435) and the Fundamental Research Funds for the Central Universities (No. 30920021112).